\documentclass[manuscript]{emulateapj}
\usepackage{amsmath}
\usepackage{natbib}
\usepackage{hyperref}

\newcommand{\mail}{liliray@pa.msu.edu}
\newcommand{\flux}{\,erg\,cm$^{-2}$\,s$^{-1}$}
\newcommand{\lum}{\,erg\,s$^{-1}$}

\newcommand{\cm}{\,cm$^{-2}$}
\newcommand{\nh}{$N_\mathrm{H}$}

\shorttitle{}
\shortauthors{Li et al.}

\begin{document}
\title{\textit{Swift}, XMM-\textit{Newton}, and \textit{NuSTAR} observations of PSR J2032+4127/MT91 213}
\author{
K. L. Li\altaffilmark{1}, A. K. H. Kong\altaffilmark{2,3}, P. H. T. Tam\altaffilmark{4}, X. Hou\altaffilmark{5,6,7}, J. Takata\altaffilmark{8}, and C. Y. Hui\altaffilmark{9}
}

\altaffiltext{1}{Department of Physics and Astronomy, Michigan State University, East Lansing, MI 48824, USA; \href{mailto:\mail}{\mail}}
\altaffiltext{2}{Astrophysics, Department of Physics, University of Oxford, Keble Road, Oxford OX1 3RH, UK}
\altaffiltext{3}{Institute of Astronomy and Department of Physics, National Tsing Hua University, Hsinchu 30013, Taiwan}
\altaffiltext{4}{School of Physics and Astronomy, Sun Yat-sen University, Zhuhai 519082, China}
\altaffiltext{5}{Yunnan Observatories, Chinese Academy of Sciences, 396 Yangfangwang, Guandu District, Kunming 650216, P. R. China}
\altaffiltext{6}{Key Laboratory for the Structure and Evolution of Celestial Objects, Chinese Academy of Sciences, 396 Yangfangwang, Guandu District, Kunming 650216, P. R. China}
\altaffiltext{7}{Center for Astronomical Mega-Science, Chinese Academy of Sciences, 20A Datun Road, Chaoyang District, Beijing 100012, P. R. China}
\altaffiltext{8}{Institute of Particle Physics and Astronomy, Huazhong University of Science and Technology, China}
\altaffiltext{9}{Department of Astronomy and Space Science, Chungnam National University, Daejeon, Republic of Korea}

\begin{abstract}
We report our recent \textit{Swift}, \textit{NuSTAR}, and XMM-\textit{Newton} X-ray and \textit{Lijiang} optical observations on PSR J2032+4127/MT91 213, the $\gamma$-ray binary candidate with a period of 45--50 years. The coming periastron of the system was predicted to be in November 2017, around which high-energy flares from keV to TeV are expected. Recent studies with \textit{Chandra} and \textit{Swift} X-ray observations taken in 2015/16 showed that its X-ray emission has been brighter by a factors of $\sim$10 than that before 2013, probably revealing some on-going activities between the pulsar wind and the stellar wind. 
Our new \textit{Swift}/XRT lightcurve shows no strong evidence of a single vigorous brightening trend, but rather several strong X-ray flares on weekly to monthly timescales with a slowly brightening baseline, namely the \textit{low state}. 
The \textit{NuSTAR} and XMM-\textit{Newton} observations taken during the flaring and the low states, respectively, show a denser environment and a softer power-law index during the flaring state, implying that the pulsar wind interacted with stronger stellar winds of the companion to produce the flares. These precursors would be crucial in studying the predicted giant outburst from this extreme $\gamma$-ray binary during the periastron passage in late 2017. \\
\end{abstract}
\keywords{X-rays: binaries --- pulsars: individual (PSR~J2032+4127) --- stars: individual (MT91~213) --- stars: winds, outflows}

\section{Introduction}
Gamma-ray binaries are a subclass of high-mass X-ray binaries (HMXBs) that harbours a compact object (neutron star or stellar-mass black hole) and a massive O or Be companion emitting modulated $\gamma$-ray emission at GeV/MeV and even TeV energies \citep{2013A&ARv..21...64D}. For those with a highly eccentric orbit of $e\gtrsim0.8$, the periastron passage of the compact object (probably neutron star in these cases as pulsar wind is usually required in modelling; see, e.g., \citealt{2013A&ARv..21...64D,2015ApJ...798L..26T} and the references therein) through the stellar wind and/or the Be circumstellar disc (if present) can trigger extraordinary flares seen from radio to TeV $\gamma$-rays (e.g., PSR~B1259$-$63/LS~2883; \citealt{2004MNRAS.351..599W,2005A&A...442....1A,2011ApJ...732L..10M,2011ApJ...736L..10T,2011ApJ...736L..11A,2015ApJ...798L..26T,2015ApJ...811...68C,2015MNRAS.454.1358C}, and HESS~J0632+057/MWC~148; \citealt{2009ApJ...690L.101H,2009ApJ...698L..94A,2009MNRAS.399..317S,2011ApJ...737L..11B,2012MNRAS.421.1103C}). \\

PSR J2032+4127/MT91~213 (J2032 hereafter) is a strong $\gamma$-ray binary candidate with a high eccentricity. It was first discovered as a $\gamma$-ray and radio emitting pulsar with the \textit{Fermi Large Area Telescope} (LAT; \citealt{2009Sci...325..840A}) and the NRAO \textit{Green Bank Telescope} (GBT; \citealt{2009ApJ...705....1C}), respectively, and later identified as a binary system with further $\gamma$-ray and radio observations \citep{2015MNRAS.451..581L}. 
While J2032 was initially thought to be a binary with a long period of $\sim$20 years, \cite{2017MNRAS.464.1211H} refined the binary model and suggested an even longer period of 45--50 years. 
According to their timing solutions, a strong radio/$\gamma$-ray pulsation at $P_s=6.98$~Hz with a strong spin-down rate of $\sim6\times10^{-13}$~s$^{-2}$ (spin-down luminosity: $\dot{E}\sim10^{35}$\lum) was detected, showing that it is a young pulsar of a characteristic age of $\sim200$ kyr. A $V=11.95$~mag Be star, MT91~213 (a member of the Cyg OB2 stellar association; about 1.5 kpc from us) is found at the inferred pulsar's position as the high-mass companion of the pulsar. The best-fit ephemeris shows that the next periastron of the binary will be in late 2017 (i.e., MJD 58069 in the \textit{Model~2} of \citealt{2017MNRAS.464.1211H}). \\

\begin{figure*}
\centering
\includegraphics[width=0.9\textwidth]{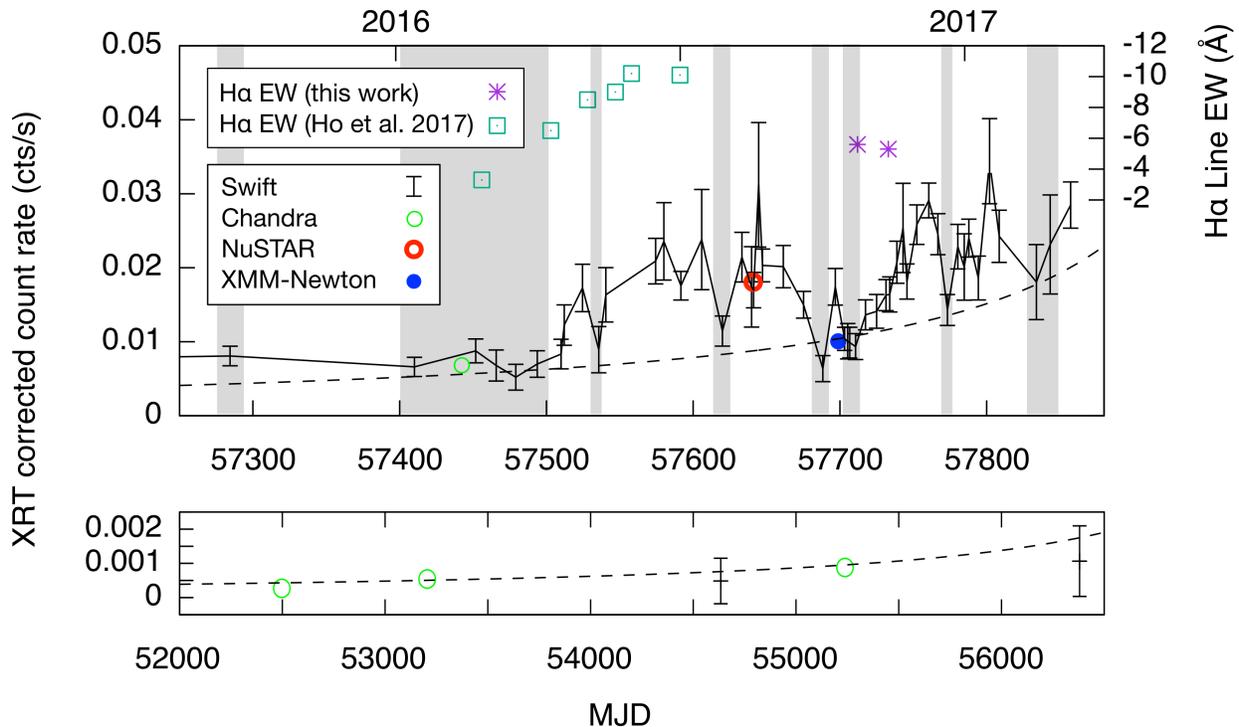}
\caption{\textit{Upper Panel} shows (i) the \textit{Swift}/XRT lightcurve (0.3--10 keV) of J2032 (black bars); and (ii) the equivalent widths (EWs) of the H$\alpha$ emission lines measured by the MDM 1.3/2.4-m and Liverpool 2-m telescopes in \cite{2017MNRAS.464.1211H} (green squares) and the \textit{Lijiang} 2.4-m telescope in this work (purple asterisks). In addition, the X-ray intensities measured by the \textit{Chandra} \citep{2017MNRAS.464.1211H}, \textit{NuSTAR}, and XMM-\textit{Newton} observations were converted to XRT's count rates based on the best-fits power-law models and shown as circles in the plot. The shadowed regions indicate the possible periods when the X-ray source was in the low state. The dashed line represents the increasing trend of the low state (i.e., $F_X\propto t_p^{-1.2\pm0.1}$, where $t_p$ is the number of days from the periastron passage on MJD 58069; see a more detailed description in \S\ref{sec:long}). \textit{Lower Panel} plots the \textit{Chandra} \citep{2017MNRAS.464.1211H} and \textit{Swift}/XRT data taken before 2013 March, with the same dashed trend line as shown in the upper panel. 
\\
}
\label{fig:swlc}
\end{figure*}

\cite{2017MNRAS.464.1211H} found an X-ray counterpart of J2032 with \textit{Chandra} and \textit{Swift}/XRT, which was faint (i.e., $F_X=(1$--$5)\times10^{-14}$\flux) before 2013, but was $\sim10$ times brighter (i.e., $F_X\approx3\times10^{-13}$\flux) after 2015. 
This extraordinary X-ray brightening strongly indicates an intimate interaction between the pulsar and stellar winds (see \citealt{2017ApJ...836..241T} for a detailed modelling). Since the brightening, a rapidly increasing trend seemingly appears in the \textit{Swift}/XRT lightcurve from 2015 September to mid-2016 \citep{2017MNRAS.464.1211H,2017ApJ...836..241T}, which is reminiscent of PSR~B1259$-$63/LS~2883 just before the disc passage (see, e.g., \citealt{2015ApJ...798L..26T,2015MNRAS.454.1358C}). \\

In this letter, we report our recent \textit{Swift}, \textit{NuSTAR}, and XMM-\textit{Newton} X-ray and \textit{Lijiang} optical observations of J2032 and clarify the current status of the system based on the results. \\

\section{The 2016 \textit{Chandra} Observation}
We re-analysed the 4.9~ksec \textit{Chandra} observation taken on 2016 February 24 (ID: 18788). While it has been well studied by \cite{2017MNRAS.464.1211H} for J2032, we focus on the three bright nearby X-ray sources (Cyg~OB2~4, MT91~221, and CXOU J203213.5+412711), which are just marginality resolved by XMM-\textit{Newton} and \textit{Swift}, and unresolved by \textit{NuSTAR}. 
Latest spectral information of these sources from the \textit{Chandra} data is extremely useful to eliminate their undesired contributions in our data. 
\\

The \texttt{CIAO} (v4.7.2) task \texttt{specextract} was used to extract the spectra with circular source regions of $r=1\farcs5$ and source-free background regions of $r=10\arcsec$. The sources can be described by an absorbed power-law or an absorbed thermal \texttt{mekal} model (the best-fit parameters are listed in Table \ref{tab:chan_spec}). These models will be included in the \textit{NuSTAR} and XMM-\textit{Newton} spectral fits (with frozen parameters) to subtract the field sources' contributions. For J2032, we used and discussed the results presented in \cite{2017MNRAS.464.1211H} throughout this work. \\

\begin{table}
\centering 
\caption{X-ray Spectral Properties of the three field sources from the \textit{Chandra} data taken on 2016 February 24}
\begin{tabular}{@{}lccc}
\hline
Source & \nh\ ($10^{22}$\cm) & $\Gamma$ or $kT$ & Norm.\footnote{See \url{https://heasarc.gsfc.nasa.gov/xanadu/xspec/manual/Models.html} for the definitions. } \\
\hline
\hline
Cyg~OB2~4\footnote{The \texttt{XSPEC} model \texttt{mekal} was used. } & 1.0 & 0.5~keV & $1.0\times10^{-4}$ \\
MT91~221 & 0.5 & 2.4 & $1.1\times10^{-5}$\\
CXOU J203213.5 & 0.8 & 2.4 & $8.6\times10^{-6}$\\
\hline
\end{tabular}
\label{tab:chan_spec}
\end{table}

\begin{figure*}
\centering
\includegraphics[width=0.49\textwidth]{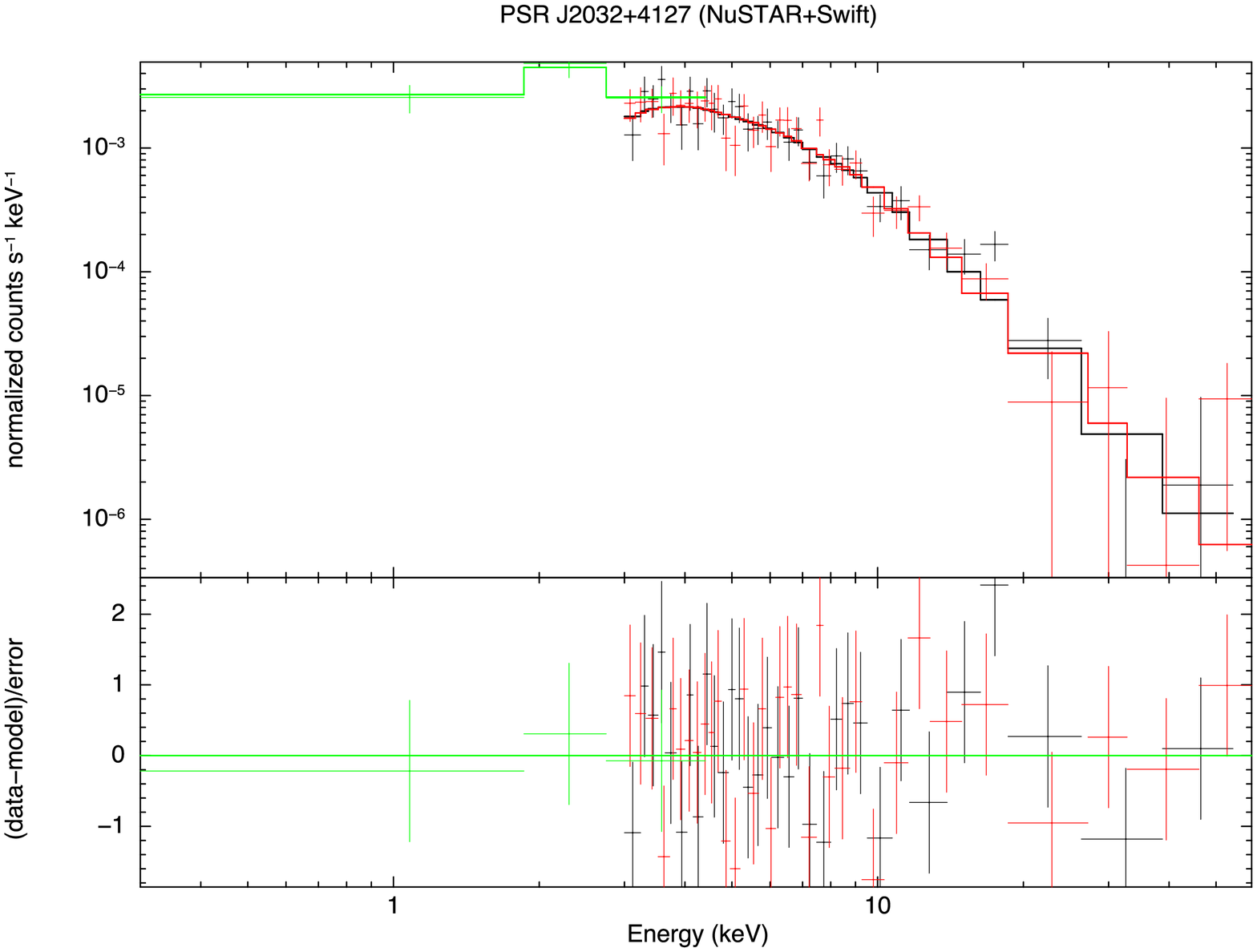}
\includegraphics[width=0.49\textwidth]{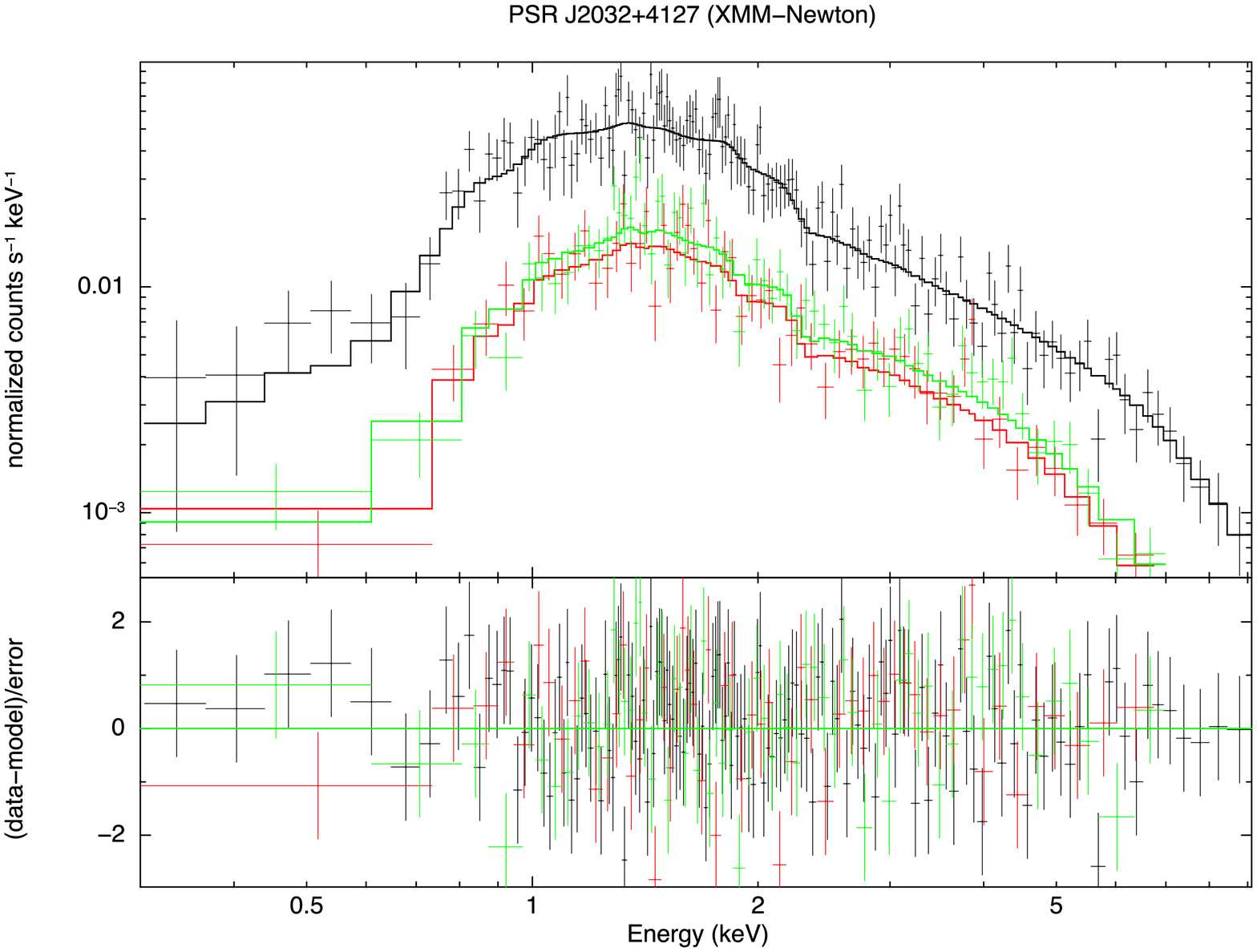}
\caption{\textit{Left}: Joint \textit{NuSTAR} (black: FPMA and red: FPMB) and \textit{Swift}/XRT (green) X-ray spectrum taken on 2016 September 9--10, with an absorbed power-law of $\Gamma=2.7\pm0.2$, \nh\ $=2.5^{+1.3}_{-0.9}\times10^{22}$\cm, and $F_\mathrm{3-78keV}=1.25^{+0.14}_{-0.13}\times10^{-12}$\flux\ (unabsorbed). 
\textit{Right}: XMM-\textit{Newton} X-ray spectrum (black: PN, red: MOS1, and green: MOS2) taken on 2016 November 6--7, with an absorbed power-law of $\Gamma=1.9\pm0.1$, \nh\ $=0.70^{+0.08}_{-0.07}\times10^{22}$\cm, and $F_\mathrm{0.3-10keV}=0.87^{+0.07}_{-0.06}\times10^{-12}$\flux\ (unabsorbed). \\
}
\label{fig:nu_spec}
\end{figure*}

\section{\textit{Swift}/XRT Observations}
\label{sec:sw}
In March 2016, we launched a bi-weekly monitoring campaign to follow up the X-ray brightening seen by \textit{Swift} and \textit{Chandra} \citep{2017MNRAS.464.1211H}. 
We once switched the observing cadence to one week from December 2016 to February 2017. But we changed it back to the two-week cadence in March, which is the best for the study. 
We also note that there is another \textit{Swift} program on J2032, probably with a longer cadence (PI: Coe). 
Some of the \textit{Swift} observations (i.e., data taken before 2016 September) have been reported in \cite{2017MNRAS.464.1211H} and \cite{2017ApJ...836..241T} and we extend the analysis with all the XRT observations taken before 2017 April 14 in this work. \\

The exposures are ranging from $<1$~ksec to 5~ksec. Most of them are useful in building a long-term X-ray lightcurve, but the data qualities are still insufficient for meaningful spectral analyses and such analyses are therefore skipped in this work. 
For the XRT lightcurve extraction, we used the \textit{Swift}'s on-line analysis tool\footnote{\url{http://www.swift.ac.uk/user_objects/}} \citep{2007A&A...469..379E,2009MNRAS.397.1177E} to take a good care of the bad pixels, vignetting, and point spread function (PSF) corrections of the data. All parameters were left at program default values with the option \textit{binning by observation} chosen. 
Figure~\ref{fig:swlc} shows the XRT lightcurve after (i) removing the bad data (i.e., upper limits due to extremely short exposures, some data bins with S/N$<3$, and a fake detection in 2006 due to a noisy background), (ii) re-binning the data points taken within 24 hours, and (iii) subtracting the expected contributions from the three bright X-ray sources (i.e., $1.7\times10^{-3}$ cts/s, estimated by \texttt{PIMMS} with the parameters in Table \ref{tab:chan_spec}). \\

As previously mentioned, the spectral information of the XRT data is bad. Given that J2032 showed strong spectral variability (cf. Table \ref{tab:xray_spec}), we discuss the XRT lightcurve using the XRT count rate throughout the paper to avoid providing misleading information. We here give a counts-to-flux conversion factor of $9.5\times10^{-11}$ erg cm$^{-2}$ cts$^{-1}$ (absorption corrected), computed based on the best-fit model for the XMM-\textit{Newton} data (see \S\ref{sec:xmm} and Table \ref{tab:xray_spec}), for a rough reference. \\

\section{\textit{NuSTAR} Observation}
We obtained a 45~ksec (live time) \textit{NuSTAR} ToO observation to study the J2032 on 2016 September 9--10 (Figure \ref{fig:swlc}). In the \textit{NuSTAR} FPMA/FPMB images, the stray light from the HMXB Cygnus X-3, 30\arcmin away from J2032, created ghost ray patterns through single reflections (Kristin Madsen, private communication). Fortunately, the contamination, especially for energies $>5$~keV, is not too severe, and the source was clearly detected with a net count rate of $\sim0.02$ cts/s (FPMA+B). A simultaneous 4~ksec \textit{Swift} observation was also obtained to extend the analysis down to 0.3~keV. \\

The \texttt{HEAsoft} (v6.19) task \texttt{nuproducts} with the CALDB (v20160731) was used to extract spectra and lightcurves from the FPMA/FPMB observations in the default energy range of 3--78~keV (channels: 35--1909). We adopted a circular source region of radius $r=30\arcsec$, which is recommended for faint sources by the \textit{NuSTAR} team. To minimize the effect of the stray light, we selected two source-free regions of $r=30\arcsec$ at the respective positions of the source in the ghost patterns for the background extractions. \\

The spectra (together with the simultaneous \textit{Swift}/XRT spectrum extracted by the \textit{Swift}'s on-line analysis tool) can be well described ($\chi_\nu^2=53.2/64$) by an absorbed simple power-law of $\Gamma=2.7\pm0.2$, \nh\ $=2.5^{+1.3}_{-0.9}\times10^{22}$\cm, and $F_\mathrm{3-78keV}=1.25^{+0.14}_{-0.13}\times10^{-12}$\flux\ (or $F_\mathrm{0.3-10keV}=6.1^{+3.2}_{-1.9}\times10^{-12}$\flux; absorption corrected), with no obvious high-energy exponential cutoff feature (Figure~\ref{fig:nu_spec}). 
The fitting result does not significantly change, if the \textit{Swift} data is not included (Table \ref{tab:xray_spec}). An additional \texttt{mekal} thermal component with a plasma temperature of $T\approx0.5$~keV can slightly improve the joint \textit{NuSTAR}-\textit{Swift} fit by $\Delta \chi^2=2.4$. We simulated 10000 spectra based on the best-fit simple power-law and then fitted the simulated spectra with both \texttt{power+mekal} and \texttt{power} models. 46\% of the simulations improve better than $\Delta \chi^2=2.4$, indicating that improvement is not significant. For the \textit{NuSTAR} lightcurve, we binned it with 5~ksec to achieve about 100 counts per bin and no strong variability can be seen. \\

\begin{table*}
\centering 
\caption{X-ray spectral fitting results of PSR~J2032+4127/MT91~213}
\begin{tabular}{@{}ccccccccc}
\hline
Date & Instruments & ${C_1}$\footnotemark[1] & ${C_2}$\footnotemark[2] & \nh\ & $\Gamma$ & $F_{\mathrm{0.3-10keV}}$\footnotemark[3] & $\chi^2/dof$\\
(MJD)&&&& ($10^{22}$\cm) &  & ($10^{-12}$\flux) \\
 \hline
 \hline\\
2016 Sept 9--10 & \textit{NuSTAR} \& \textit{Swift} &$1.0\pm0.1$&$0.9^{+0.4}_{-0.3}$& $2.5^{+1.3}_{-0.9}$ & $2.7\pm0.2$ &$6.1^{+3.2}_{-1.9}$& 53.2/64\\

2016 Sept 9--10 & \textit{NuSTAR} &$1.0\pm0.1$& \nodata & $<5.3$ & $2.6^{+0.3}_{-0.2}$ &$5.0^{+5.6}_{-1.8}$& 52.8/62\\

2016 Nov 6--7 &XMM-\textit{Newton} &$0.91\pm0.06$&$0.95\pm0.06$& $0.70^{+0.08}_{-0.07}$ & $1.9\pm0.1$ &$0.87^{+0.07}_{-0.06}$& 278.9/272\\
\hline
\end{tabular}
\footnotetext{The cross-calibration factor of FPMB (or MOS1) w.r.t. FPMA (or PN). }
\footnotetext{The cross-calibration factor of XRT (or MOS2) w.r.t. FPMA (or PN). }
\footnotetext{The fluxes have been absorption corrected. }
\label{tab:xray_spec}
\end{table*}

\section{XMM-\textit{Newton} Observation}
\label{sec:xmm}
A 43~ksec XMM-\textit{Newton} ToO observation operated under the prime full window mode with the medium optical blocking filter was obtained on 2016 November 6--7 (Figure~\ref{fig:swlc}). 
Following the analysis threads in the XMM-\textit{Newton} Science Operation centre\footnote{\url{https://www.cosmos.esa.int/web/xmm-newton/sas-threads}}, we used the metatask \texttt{xmmextractor} in \texttt{SAS} (v15.0.0) to extract the scientific products from the raw data in the Observation Data File (ODF). \\

The live times were 35~ksec and 41~ksec for PN and MOS1/2, respectively. After filtering the high background periods, usable live times are reduced to 27~ksec (PN), 39~kec (MOS1), and 38~kec (MOS2). J2032 was detected in all EPIC cameras with net count rates of 0.1~ cts/s for PN and 0.03~cts/s for each MOS in 0.3--10~keV. Similar to the \textit{NuSTAR} lightcurve, no hourly variability can be seen in the PN (1~ksec binned) and MOS1+2 (1.5~ksec binned) lightcurves. We fitted an absorbed simple power-law model to the spectra and found the best-fit parameters of $\Gamma=1.9\pm0.1$, \nh\ $=0.70^{+0.08}_{-0.07}\times10^{22}$\cm, and $F_\mathrm{0.3-10keV}=0.87^{+0.07}_{-0.06}\times10^{-12}$\flux\ (absorption corrected; $\chi_\nu^2=278.9/272$) (Figure \ref{fig:nu_spec}), which are all very different from that of the \textit{NuSTAR}+\textit{Swift} spectral fit (Table \ref{tab:xray_spec} and Figure \ref{fig:con}). We also tried to add a \texttt{mekal} component to improve the fit, but the reduced $\chi$ was found to be even higher. All the best-fit parameters (including the \textit{NuSTAR+Swift}'s) are shown in Table \ref{tab:xray_spec}. \\

\section{The \textit{Lijiang} 2.4-m Observations}
To study the evolving H$\alpha$ emission line from the circumstellar disc of MT91~213 \citep{2017MNRAS.464.1211H}, two $120$~sec and $180$~sec spectra were taken with the \textit{Yunnan Faint Object Spectrograph and Camera} (YFOSC) on the \textit{Lijiang} 2.4-m telescope on 2016 November 20 and December 11, respectively. 
The spectral resolutions are medium with Grism 15 (183 nm/mm) \& Slit 3 ($1\farcs0$) on December 11, and Grism 14 (92 nm/mm) \& Slit 3 ($1\farcs8$) on November 20. After the standard data reduction processes with \texttt{IRAF}, the H$\alpha$ emission line was clearly detected in both datasets, although the double-peaked line profile \citep{2017MNRAS.464.1211H} is unresolved. Using the \texttt{eqwidth} task in the \texttt{rvsao} package, we computed the equivalent widths (EW) of the H$\alpha$ emission lines to be $-5.6$\AA\ and $-5.3$\AA\ on November 20 and December 11, respectively (Figure~\ref{fig:swlc}). \\

\section{Discussion}
\label{sec:dis}
Before 2013 March, the X-ray source was marginally detected by XRT at $C_\mathrm{XRT}\sim0.001$ cts/s. 
In 2015 September, it had brightened to $C_\mathrm{XRT}\approx0.008$ cts/s after a 2.5-year observing gap. 
The X-ray emission then increased more rapidly from $C_\mathrm{XRT}\approx0.007$ cts/s to $\approx0.024$ cts/s in 2016 April--July \citep{2017MNRAS.464.1211H,2017ApJ...836..241T}. 
While a continuous increase was theoretically expected (see, e.g., \citealt{2017ApJ...836..241T}), the X-ray emission returned back to $C_\mathrm{XRT}\approx0.006$ cts/s in three months, confirmed by the XMM-\textit{Newton} observation (Figure \ref{fig:swlc}). The flux was increasing again afterwards, but a few declines are again shown later (Figure \ref{fig:swlc}). 
More than one type of variation should be involved to result in the complexity seen in the X-ray lightcurve. \\

\subsection{The Long-Term Variability}
\label{sec:long}
By taking a closer look of the \textit{Swift}/XRT lightcurve, one can easily identify some local flux minima and the most obvious ones are indicated by the shadowed regions in Figure \ref{fig:swlc}. 
We tried to fit these XRT minima in the shadow and the quiescent fluxes measured before 2013 (including the three \textit{Chandra} measurement taken in 2002--2010; \citealt{2017MNRAS.464.1211H}) to a simple power-law, and the data can be well connected with $F_X\propto t_p^{-1.2\pm0.1}$ (Figure \ref{fig:swlc}), where $t_p$ is the number of days from the periastron passage (i.e., MJD 58069; \citealt{2017MNRAS.464.1211H}). 
Apparently, these low flux intervals could belong to the same emission state, which will be called the \textit{low state} in the following discussion. \\

The momentum ratio of the stellar wind to the pulsar wind is one of the major factors to determine the X-ray luminosity of the wind-wind interacting shock in a $\gamma$-ray binary. The dependence can be even higher under a consideration of a non-constant magnetization of the shock along the distance from the pulsar \citep{2017ApJ...836..241T}. 
In J2032, the slowly brightening low state are likely the consequence as the pulsar approaches the Be star and interacts with the stronger stellar wind. 
Because of the current large distance between the pulsar and the Be star, the rate of X-ray flux increase would be slow. However, this is still sufficient to develop the two distinct flux levels before/after the 2.5-year observing gap in 2013--2015, as observed by \textit{Swift}/XRT and \textit{Chandra}. \\

It is worth noting that the XMM-\textit{Newton} data was taken during the low state. The best-fit hydrogen column density (i.e., \nh\ $=7\times10^{21}$\cm) is well consistent with the foreground value estimated by the optical color
excess of MT91~213 (i.e., \nh\ $=7.7\times10^{21}$\cm; \citealt{2009ApJ...705....1C,2017MNRAS.464.1211H}). In addition, the best-fit photon index (i.e., $\Gamma=1.9$) is very close to that of those \textit{Chandra} observations taken before 2016 (i.e., $\Gamma=2$ with the foreground \nh; \citealt{2017MNRAS.464.1211H}), supporting our suggestion that the source was in the same low state during the \textit{Chandra} observations. \\

\begin{figure}
\centering
\includegraphics[width=0.45\textwidth]{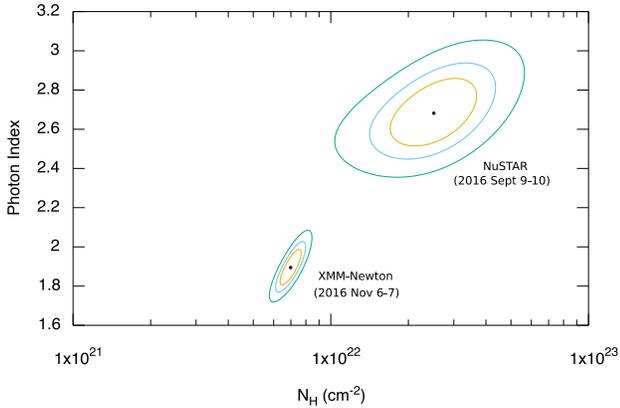}
\caption{In the hydrogen column density (\nh) and photon index plane, significant differences in both axes are seen between the contours (68\%, 90\%, and 99\%) for the \textit{NuSTAR} and XMM-\textit{Newton} fits. 
}
\label{fig:con}
\end{figure}

\subsection{The Short Variability}
Besides the possible long-term brightening trend, multiple flares on weekly to monthly timescales are obviously present in the XRT lightcurve (Figure \ref{fig:swlc}). Our \textit{NuSTAR} observation provides a good X-ray spectroscopic study on one of these flares. Comparing with the low state spectrum taken by XMM-\textit{Newton}, the \textit{NuSTAR} spectrum is significantly softer in photon index with a heavier \nh\ absorption (Figure \ref{fig:con}). 
This high \nh\ strongly implies a denser medium around the pulsar during the flare, probably caused by an occasional strong wind from the Be star. 
Using the binary orbit presented in \cite{2017MNRAS.464.1211H} and the mass-loss rate of $\dot{m}=4\pi\,r^2v_w\rho_w$ (where $r$ is the distance from the star, $v_w\sim1000$~km/s is the wind speed, and $\rho_w$ is the density of the wind at $r$) for a steady and spherically symmetric wind, we integrated the density along the line-of-sight and found that $\dot{m}\sim10^{-5}$--$10^{-4}M_\sun$ yr$^{-1}$ is required to accumulate the intrinsic \nh\ to $\sim 10^{22}$\cm. This inferred rate is several orders of magnitude higher than the typical value for B type stars \citep{2014A&A...564A..70K}, suggesting that the wind is likely compact and clumpy (i.e., the pulsar was hitting compact wind clumps, instead of a homogeneous wind). 
When impacting the pulsar, this strong clumpy wind probably pushed the shock toward the pulsar side to cause a stronger magnetic field at the emission region \citep{2017ApJ...836..241T}. 
In this case, \textit{NuSTAR} might observe the emissions from the particles in both the slow and fast cooling regimes in the flaring state, while only the emission from the slow cooling regime was observed by XMM-\textit{Newton} in the low state, possibly explaining the observed divergence in photon index. \\

In the H$\alpha$ line study of \cite{2017MNRAS.464.1211H}, the circumstellar disc of the Be star was expanding from $R_\mathrm{disc}\approx0.2$ to 0.4 AU (i.e. EW: from $-3.3$ to $-10.2$\AA) in 3--4 months during the first few X-ray flares (see also Figure \ref{fig:swlc}). 
Our \textit{Lijiang} spectra indicate that the circumstellar disc shrank back to $R_\mathrm{disc}\approx0.3$ AU (converted from EW using the equation in \citealt{1989Ap&SS.161...61H}) a few months later while the X-ray source was likely in the low state. 
We suspect that the disc expansion is an indication of the hypothetical strong clumpy wind to trigger the observed flares. On the other hand, the activity of the circumstellar disc could be induced by the approaching pulsar as shown in PSR~B1259$-$63/LS~2883 \citep{2014MNRAS.439..432C,2015MNRAS.454.1358C}, although the separation between the pulsar and the Be star in J2032 was much larger. \\


Finally, we note that PSR~B1259$-$63/LS~2883 did not show such pre-periastron-passage flares previously (see, e.g., \citealt{2006MNRAS.367.1201C}). However, when PSR~B1259$-$63 entered the circumstellar disc in 2004, the X-ray emission, \nh, and the photon index were all increasing \citep{2006MNRAS.367.1201C}. In J2032, a very similar spectral change is seen when transiting from the low state to the flaring state (Table \ref{tab:xray_spec} and Figure \ref{fig:con}), although the photon indexes of PSR~B1259$-$63/LS~2883 are generally harder (i.e., increased from $\Gamma=1.2$ to 1.8) than that of J2032 (i.e., from $\Gamma=1.9$ to 2.7). It would be intriguing to ask whether this is a common feature in $\gamma$-ray binaries when the pulsars entering from a lighter medium to a denser medium. \\

\section{Conclusion}
With the \textit{NuSTAR} and XMM-\textit{Newton} X-ray observations, we identify two very different spectral states, namely the low state (i.e., low X-ray flux and \nh\ with a hard spectrum) and the flaring state (i.e., high X-ray flux and \nh\ with a soft spectrum). The \textit{Swift}/XRT lightcurve suggests that the low state has been slowly evolving, possibly following $F_X\propto t_p^{-1.2}$, while the flares are likely on weekly to monthly timescales. In addition, these flares could be correlated to the size of the circumstellar disc of MT91~213, indicated by the H$\alpha$ emission line studies (see also \citealt{2017MNRAS.464.1211H}). The physical origin of these flares and the implication of the slowly brightening low state are still not entirely clear. Hopefully, continuous multi-wavelength monitoring observations (e.g., from \textit{Swift} and \textit{Fermi}) will be useful in studying these flares as well as any pre-periastron activities before the periastron passage in late 2017. \\

\begin{acknowledgements}
We thank (i) Neil Gehrels and Brad Cenko for approving our \textit{Swift} monitoring campaign, (ii) Fiona Harrison for approving the \textit{NuSTAR} DDT request and Kristin Madsen for the technical support, and (iii) Norbert Schartel for approving the XMM-\textit{Newton} DDT request and Rosario Gonzalez-Riestra for scheduling the observation. 
We acknowledge the support of the staff of the \textit{Lijiang} 2.4m telescope. Funding for the telescope has been provided by Chinese Academy of Sciences and the People’s Government of Yunnan Province. The scientific results reported in this article are based in part on data obtained from the \textit{Chandra} Data Archive. 
AKHK is supported by the Ministry of Science and Technology of Taiwan through grant 105-2112-M-007-033-MY2, 105-2119-M-007-028-MY3, and 106-2918-I-007-005. XH is supported by National Natural Science Foundation of China through grant 11503078. 
PHT is supported by the National Science Foundation of China (NSFC) through grants 11633007 and 11661161010. 
JT is supported by NSFC grants of Chinese Government under 11573010 and U1631103. 
CYH is supported by the National Research Foundation of Korea through grants 2014R1A1A2058590 and 2016R1A5A1013277. 
\\
\end{acknowledgements}
\textit{Facilities}: \facility{Swift, NuSTAR, XMM, YAO:2.4m, and CXO}


\end{document}